# Quantization of Time in Dynamic Barrier Tunnelling

Sydney G. Davison[1,2,4] and Timothy S. Davison [3,5]

[1] *Department of Applied Mathematics, University of Waterloo, Waterloo, ON, Canada, N2L 3G1*
[2] *Department of Physics, University of Waterloo, Waterloo, ON, Canada, N2L 3G1*
[3] *Centre for Cancer Research and Cell Biology, Queen's University, Belfast, UK, BT9 7BL*
[4] Corresponding author: sgdaviso@uwaterloo.ca;   [5] t.davison@qub.ac.uk

(Dated: April 9, 2016)

## 1. Abstract

In the Büttiker-Landauer perturbation approach to electron tunnelling, through a time-modulated rectilinear potential barrier, the Tien-Gordon identity was invoked, together with its *infinite* energy spectrum. Here, an *exact* treatment is presented which is based on the temporal wave-function matching procedure, that led to a *finite* energy spectrum. In seeking the condition governing the time evolution of the tunnelling process, Euler's formula provided the crucial ingredient for *time quantization*, which discretised the continuous time in the oscillating barrier's potential and energy harmonic equations. As a result, a finite system of inelastic scattering channels was created. When an electron entered the elastic channel, it was scattered, instantaneously, into finite neighbouring energy-level scattering channels, by absorption (emission) of photon energy from (to) the oscillating field, during the transit period across the dynamic barrier. The absorption and emission times of barrier traversal, $T_+$ and $T_-$, respectively, were derived for the low and high frequency regimes of the barrier oscillations. Calculations revealed that in the low (high) frequency situation $T_+ = T_-$ ($T_+ < T_-$).

## 2. Introduction

Although solutions of the time-dependent Schrödinger equations (TDSE) for a time-periodic potential appeared in the 1960s [1,2], almost two decades elapsed before the seminal work of Büttiker and Landauer (BL) [3], on the electron transmission properties of time-modulated potential barriers, sparked renewed interest in time-periodic potential problems. In the BL model, an electron of energy $E_0$ is incident on a static rectangular barrier, whose height $V_0$ is time-modulated by a harmonically oscillating perturbing potential of small amplitude $V_1$ and frequency $\omega$. During its passage through such a barrier, the electron is scattered into *sideband* energy levels, created by the time-modulation process. The *infinite* energy spectrum, so generated, is described by the expression $E_n = E_0 + n\hbar\omega$, where $n = 0$, $\pm 1, \pm 2, \ldots$, $E_0$ is the energy of the *centre-band* (elastic channel) and $E_n$ denotes the energy of the nth sideband (inelastic scattering channel). After deriving the well-known relation for the static-barrier centre-band transmission probability $T(E_0)$, BL performed a first-order perturbation ($V_1/\hbar\omega$) calculation to obtain the corresponding probabilities $T(E_{\pm 1})$ for the first sideband energies of a time=modulation (*dynamic*) barrier. In a further discussion, they introduced the controversial concept of a time of traversal of a barrier [3-7]. Among the authors [8], who questioned the BL concept of the barrier time of traversal, the most severe critic was Truscott, who expounded a novel coordinate-transformation method, which eliminated the temporal field, as a key to an exact solution of the TDSE. The approach espoused here, addresses the time-modulated barrier issue, by proposing the traditional treatment of wave-function matching, which provides an alternative exact solution for the barrier transit time (see section 4).

In contrast to the BL work on dynamic-barrier transmission, Wagner [9] focused his attention on the transmission probability of a driven *quantum-well*, whose base was subjected to a harmonic driving potential. In matching such a well between two static barriers, it was found that, unlike the oscillating-barrier case, sideband intensities showed no monotonic increase with $V_1$ and decreased with $|n|$. Moreover, all sideband transmission probabilities experienced a simultaneous quenching for certain characteristic ($V_1/\hbar\omega$) values. Subsequently, Wagner [10] solved analytically the TDSE for the lowest Floquet state [2] in a single quantum-well driven by a very strong laser field. Floquet states were also used by Grossman et al [11] to study a harmonically driven symmetric double-well potential.

At this juncture, a comprehensive introduction to the basic formulation of one-dimensional scattering by time-periodic short-ranged potentials was provided by Saraga and Sassoli de Bianchi [12]. They emphasised the connection between the time-dependent approach and the quasi-stationary one. The general Born expansion was applied to the BL, model and the square-barrier oscillating in position, to calculate the full transmission probability in each case, up to the first non-vanishing correction in the time-dependent perturbation.

Next we discuss the work of Reichl and his co-workers. Firstly [13], they construct the Floquet *S*-matrix to obtain the transmission probabilities and Wigner delay times for electrons passing through a harmonically driven potential. It was found that the interaction of the electrons with the oscillating field gave rise to transmission resonances via photon emission and absorption. When oscillator-induced quasi-bound states trap electrons, they can produce electron inter-channel transitions at the resonances. The presence of an oscillating field produced an ac Stark effect. In the complex-energy plane, the Floquet quasi-bound states appeared at transmission poles. Meanwhile, lifetimes derived from these poles turned out to be of the same order of magnitude as the corresponding Wigner delay times. Secondly [14], the continued-fraction method was used, within the Floquet theory framework, to derived the exact expression for the transmission amplitude of a particle moving through a harmonically driven *δ*-function potential. After a detailed analysis of the zeros and poles in the transmission amplitude, the existence of non-resonant "bands" in the amplitude was shown to be a function of the potential strength and the driving frequency.

Since the BL [3] model invokes the Tien-Gordon (TG) [1] identity, it is instructive to briefly outline the method by which they obtained it. The derivation begins by expressing the oscillating barrier's temporal wave-function in terms of an infinite Fourier series expansion, which, when inserted in the temporal part of the TDSE, leads to the expansion coefficients being expressed as nth order Bessel functions of the first-kind in the identity

$$\exp[-i(V_1/\alpha)\sin\omega t] = \sum_{n=-\infty}^{\infty} J_n(V_1/\alpha)e^{-in\omega t},$$

which is accompanied by the infinite energy spectrum

$$E_n = E_0 \pm n\alpha,$$

where $\alpha\,(=\hbar\omega)$ is the equal spacing of photon energy and $n = 0, 1, 2, \ldots, \infty$. Mathematically, such an outcome is certainly valid, and thus, made the BL model a useful tool by its adoption.

However, there arises the crucial question of the wave-function continuity, which requires the necessary wave-function matching procedure at spatial boundaries. On pursuing this course here, one encounters the essential *quantum condition* that leads to the concept of the *quantization of time*. As will be seen, this finding creates a new perspective of electron tunnelling through time-modulated potential barriers, and gives rise to an *exact* method, as a basis for calculations. Unfortunately, as a result, the TG approach does not embody the *true essence* of the wave-function matching continuity procedure. For this reason, we adopt the well-known wave-function matching technique [15].

A paper which pays attention to the importance of continuity conditions for time-dependent wave-functions at potential discontinuities, is that due to Lefebvre [16], who used the matching of Fourier components in the transfer-matrix technique, where the TG [1] identity provides the Fourier decomposition of each Gordon-Volkov wave. There follows a numerical analysis of tests of convergence for two matching procedures, employed in calculating a rectangular barrier's transmission and transition probabilities for absorption and emission of a quantum, as pertains to the BL time of barrier traversal. As will become apparent, these findings resemble certain aspects of the present work. In particular, the possible accord between the *equality* of transmission probabilities of absorption and emission of a quantum, in the low-frequency regime, with the same *equality*, as found here, between the corresponding *quantal times* of barrier traversal.

The purpose of the present paper is to describe an *exact*, rather than perturbative, method of treating the BL model for an electron tunnelling through a time-periodic modulated rectangular potential barrier. This is achieved by *fully* exploiting the matching conditions on the temporal-parts of the wave-functions at the spatial boundaries of the time-periodic barrier. In doing so, we find that the time-dependent interaction term in the Hamiltonian does *not* change the spatial distributions of the wave-functions, but merely modifies the energy of the electron [1]. Indeed, in this way, a physically appealing *finite* discrete energy spectrum is obtained, as opposed to the infinite BL one. Under these circumstances, problems involving Hamiltonians periodic in time may be solved by methods applicable to time-independent

Hamiltonians [2], as is illustrated in our calculation of the *transmission probability* of a rectangular dynamic barrier. Finally, the quantized times of traversal of the barrier are obtained for absorption and emission of photons in the cases of low- and high-frequency regimes. Discussion of the results include comparisons with the findings of BL and Lefebvre.

### 3. Dynamic Barrier Tunnelling

On entering a dynamic-potential barrier via the elastic channel, an electron interacts with the oscillating potential field of the barrier, and is inelastically scattered into one of the sideband energy levels, created by the time-modulations. Because the second-order TDSE for a dynamic-barrier can be separated into spatial and temporal parts, it is convenient, when discussing the complete picture of the electron's tunnelling process, to treat the static and dynamic barrier's aspects separately.

### A. Static Barrier

Since this topic is found in most quantum mechanics textbooks [e.g. ref [15]], only a brief treatment is presented here. As can be seen in Figure 1(a), the static rectangular barrier of height $V_0$ and width $b$ occupies region II, which lies between the points $x = \pm b/2$, and separates the infinite free-space regions I and III, which are of zero potential. An electron of energy $E_0$ ($<V_0$) is incident on the barrier from region I, and its passage through the barrier can be described in terms of the solutions of the TDSE in the three regions. These time-modulated wave-function solutions can be written as products of spatial and temporal terms, viz.,

$$\psi_1(x,t) = (e^{ik_0 x} + A_0^- e^{-ik_0 x})e^{-iE_0 t/\hbar}, \tag{1}$$

$$\psi_2(x,t) = (B_0^+ e^{\kappa_0 x} + B_0^- e^{-\kappa_0 x})e^{-iE_0 t/\hbar}, \tag{2}$$

$$\psi_3(x,t) = C_0^+ e^{-ik_0 x} e^{-iE_0 t/\hbar}, \tag{3}$$

where

$$k_0^{\ 2} = 2mE_0/\hbar^2, \qquad \kappa_0^{\ 2} = 2m(V_0 - E_0)/\hbar^2, \tag{4}$$

$A_0^-, B_0^\pm$ and $C_0^+$ being the *amplitudes* of the regional wave functions portrayed in Figure 1 (a). The continuity of the wave functions and their derivatives at the boundaries at $x = \pm b/2$ require that they satisfy the matching conditions:

$$\psi_1(-b/2, t) = \psi_2(-b/2, t), \tag{5}$$

$$\psi_1'(-b/2, t) = \psi_2'(-b/2, t), \tag{6}$$

$$\psi_2(b/2, t) = \psi_3(b/2, t), \tag{7}$$

$$\psi_2'(b/2, t) = \psi_3'(b/2, t). \tag{8}$$

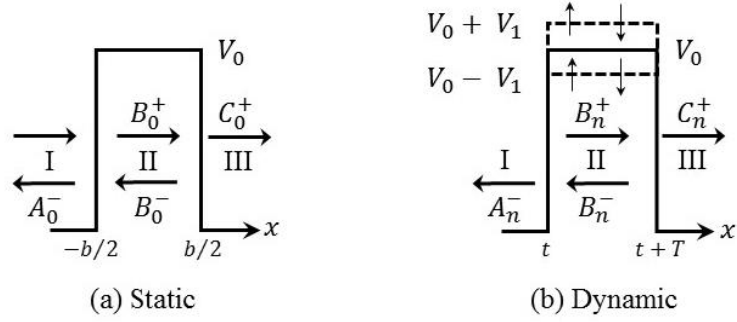

Figure 1. Wave-function amplitudes in regions I, II and III for; (a) Static barrier tunnelling, where $V_0$ is the barrier height and $b$ the width. Note that the incident amplitude in region I is unity. (b) Dynamic barrier tunnelling, where barrier height $V_0$ is subjected to time oscillations of amplitude $V_1$ between the upper (lower) limit of $V_0 + V_1$ ($V_0 - V_1$).

By inserting the wave functions (1) to (3) in these conditions, the relations for the amplitudes $A_0^-, B_0^\pm$ and $C_0^+$ can be found. In particular, we are interested in the amplitude $C_0^+$, because it provides the *transmission probability* via the equation

$$T(E_0) = |C_0^+|^2. \tag{9}$$

Since the calculations outlined above have been carried out by BL [3], we merely quote their result here, thus,

$$T(E_0) = \frac{4k_0^2 \kappa_0^2}{4k_0^2 \kappa_0^2 + (k_0^2 + \kappa_0^2)^2 \sinh^2 \kappa_0 b}. \tag{10}$$

In the special case of an almost *opaque* barrier, where $\kappa_0 b \gg 1$, the hyperbolic function in (10) is replaced by an exponential one, so that

$$T_0(E_0) = \frac{16 k_0^2 \kappa_0^2}{16 k_0^2 \kappa_0^2 + (k_0^2 + \kappa_0^2)^2 e^{2\kappa_0 b}}. \tag{11}$$

## B. Dynamic Barrier

The problem of electron transmission through a static barrier becomes considerably more difficult when the height of the barrier $V_0$ is subjected to a time-periodic potential of modulation amplitude $V_1$, say. In this situation, the time-modulated barrier potential can be represented mathematically by

$$V(t) = V_0 + V_1 \sin \omega t, \tag{12}$$

$\omega$ being the modulation frequency. For such a potential, the wave-function solution of the TDSE in the dynamic-barrier region II (Figure 1b) can be written as

6$$\psi_2(x,t) = (B_0^+ e^{\kappa_0 x} + B_0^- e^{-\kappa_0 x})e^{-i\xi(t)/\hbar}, \tag{13}$$

where

$$\kappa_0^2 = 2m(V_0 - E_0)/\hbar^2, \tag{14}$$

$$\xi(t) = \int^t E(t)dt, \tag{15}$$

$$E(t) = E_0 + V_1 \sin \omega t. \tag{16}$$

It is clear, from these equations, that their time-dependent forms do not reflect the *discrete* nature of the elastic and scattering channel energy levels created in the dynamic barrier by the time modulations. To address this problem, it is necessary to match the *temporal* parts of the free-space (1) and dynamic barrier (13) wave-functions, as indicated in (5), whereby (15) and (16) we find

$$exp(-iE_0 t/\hbar) = exp\left[\frac{-i}{\hbar}\left(E_0 t - \frac{V_1}{\omega}\cos \omega t\right)\right], \tag{17}$$

which, with the aid of Euler's formula of complex analysis gives

$$exp[i(V_1/\hbar\omega)\cos \omega t] = \exp(2\pi n i), \tag{18}$$

where $n = 0, \pm 1, \pm 2, \ldots$. Equating the exponents in (18) leads to

$$V_1 \cos \omega t_n = n\hbar\omega = n\alpha, \quad n = 0, \pm 1, \pm 2, \ldots, \tag{19}$$

which is the crucial condition giving rise to the *quantization* of the oscillating potential field in units of photon energy $\alpha\ (=\hbar\omega)$. By virtue of (19), equations (12) and (16), become the *discretized (quantized)* forms of the scattering channels' potential and energy equations, namely,

$$V_n^\pm = V_0 \pm [V_1^2 - (n\alpha)^2]^{1/2}, \tag{20}$$

$$E_n^\pm = E_0 \pm [V_1^2 - (n\alpha)^2]^{1/2}, \tag{21}$$

respectively. In light of (20), we see that the former dynamic barrier, whose height varied *continuously* with time, according to (12), has now been replaced by a *series of instantaneously static barriers*, whose heights are governed by the values of *n* in (20). In this way, an array of barrier heights is produced between the limits of $V_0 \pm V_1$ (Figure 1b). Simultaneously, the presence of *identical n*-terms in (20) and (21), ensures that, since $E_0 < V_0$, a *correspondingly sequence of discrete instantaneously stationary sideband energy levels* is created *below* the tops of the 'snapshot' barriers, between the limits of the $E_0 \pm V_1$ band edges.

A closer examination of equation (21), reveals that the *reality* of $E_n^\pm$ requires $|n\alpha| \leq V_1$. It therefore follows that the number of energy levels is given by, say,

$$N = V_1 \alpha^{-1}, \tag{22}$$

which clearly demonstrates the important role the modulation amplitude $V_1$ plays in *limiting* the extent of the energy spectrum. Noting $V_1 = N\alpha$, we see that $\alpha^{-1}$ and $\alpha$ act as *conversion* factors between $N$ and $V_1$, the former (latter) making $N$ very large ($V_1$ small). In view of (22), we can write (21) as the discrete energy of the *scattering channels*, i.e.,

$$E_n^\pm = E_N \pm (N^2 - n^2)^{1/2}\alpha, \tag{23}$$

where $E_N$ replaces $E_0$ as the more appropriate notation for the elastic-channel energy level at $n = N$. On rearranging, (23) takes the form of a *circle*, viz.,

$$(E_n^\pm - E_N)^2 + (n\alpha)^2 = (N\alpha)^2, \tag{24}$$

whose centre is located at the point $(n\alpha, E_n^\pm) = (0, E_N)$ and has a radius of $N\alpha$ $(= V_1)$ (Figure 2). As can be seen, the energy levels $E_n^\pm$ are obtained by projection on to the vertical axis, as illustrated by $E_m^+$ and $E_m^-$. We also note that the modulating potential amplitude $V_1$ limits the number of inelastic scattering-channel energy levels to $N$ *above* and *below* the centre-band energy $E_N$ of the elastic channel. Such a *finite* energy spectrum is contrary to the BL picture [3], which has an *infinite* spectrum, unrelated to the modulation amplitude $V_1$ creating it.

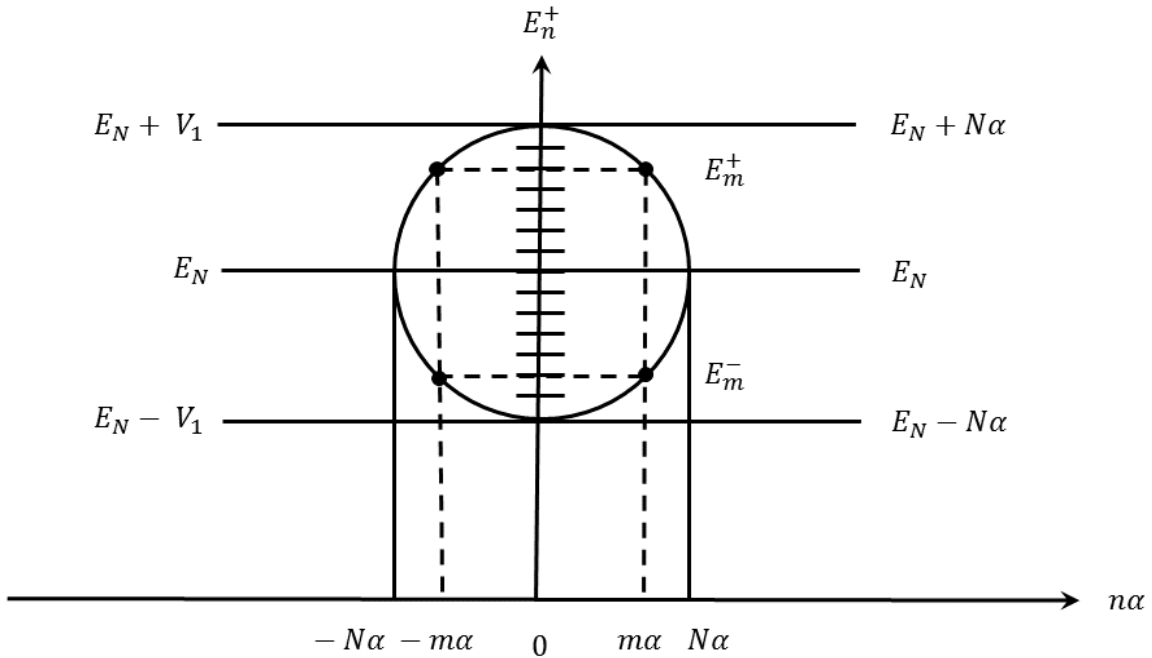

Figure 2. Energy circle of $E_n^\pm$ versus $n\alpha$, showing elastic-channel energy level $E_N$ and typical sideband scattering-channel energy levels $E_m^\pm$ lying between corresponding band-edge values $E_n \pm V_1$ and $E_n \pm N\alpha$.

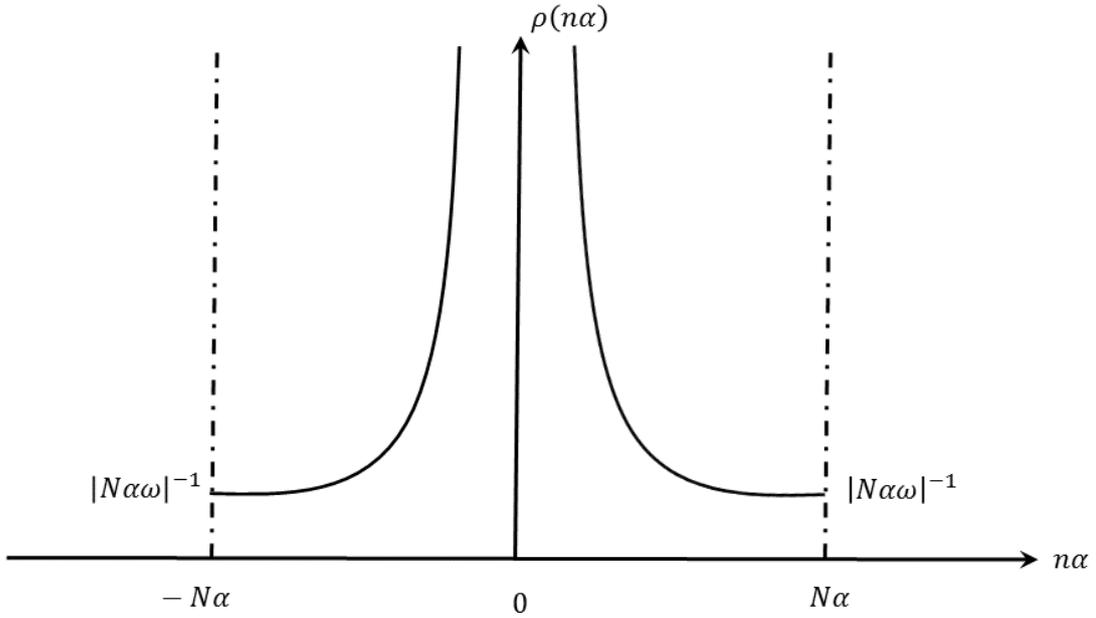

Figure 3. Density of states curves $\rho(n\alpha)$ vs $n\alpha$ with limiting values at $\rho(N\alpha) = |N\alpha\omega|^{-1}$. Energy levels are dense (sparse) at band-centre (band-edges).

In order to find the distribution of the energy levels $E_n^{\pm}$ (23) in the finite-energy band, we investigate their *density-of-states* $\rho(n\alpha)$, which is derived from (16) to give

$$\rho(t) = |dt/dE| = |\omega V_1 \cos \omega t|^{-1}. \tag{25}$$

On using (19), we obtain the *discrete* form

$$\rho(n\alpha) = |n\alpha\omega|^{-1}, \tag{26}$$

whose graph appears in Figure 3, where a pair of rectangular-hyperbolic curves are reflected in the $\rho(n\alpha)$-axis with values of $\rho(N\alpha) = |N\alpha\omega|^{-1}$ at the band-edges $\pm N\alpha$. As is apparent from Figure 3, the energy levels $E_n^{\pm}$ (23) are *dense* (*sparse*) at the band-centre (band-edges), so the energy levels are *not* equally spaced as in the BL case [3].

Having incorporated the *discrete* nature of the modulated-barrier energy levels into our calculations, we can now turn to the question of describing the dynamic-barrier wave-function. Utilizing (18) and (21), we find that (13) and (14) can be rewritten as

$$\psi_2(x,t) = \sum_n (B_n^+ e^{\kappa_n x} + B_n^- e^{-\kappa_n x}) e^{-iE_N t/\hbar}, \tag{27}$$

$$\kappa_n^2 = 2m(V_0 - E_n^{\pm})/\hbar^2, \tag{28}$$

9respectively. By the same token, the appearance of the multi-scattering channels in the dynamic barrier, means that the free-space wave-functions (1) and (3), must likewise be upgraded to reflect their presence and, hence, should be recast as

$$\psi_1(x,t) = (e^{ik_N x} + \sum_n A_n^- e^{-ik_n x})e^{-iE_N t/\hbar}, \tag{29}$$

and

$$\psi_3(x,t) = \sum_n C_n^+ e^{ik_n x} e^{-iE_N t/\hbar}, \tag{30}$$

with

$$k_n^2 = 2mE_n^\pm/\hbar^2, \tag{31}$$

via (4). The summations over n, in these equations, include both the elastic ($n = N$) and the inelastic scattering ($n \neq N$) channel energy levels. As is evident, from (27), (29) and (30), the effect of the time-dependent term in the periodic potential (12), on these regional wave-functions, manifests itself *solely* in the modification of the electron centre-band energy $E_N$, as in (23), and leaves their spatial distributions entirely *undisturbed* [1]. Such an outcome reveals that problems, involving the time-periodic potentials, may be solved by means of *time-independent* methods [2]. Consequently, since the above wave-functions apply to a dynamic barrier, mimicked by a sequence of snapshot *static* barriers, it follows that we can obtain the *transmission probability* for the *n*th channel directly from the *static* case (10), by simply rewriting it in the form

$$T_n(E_n^\pm) = \frac{4k_n^2 \kappa_n^2}{4k_n^2 \kappa_n^2 + (k_n^2 + \kappa_n^2)^2 \sinh^2 \kappa_n b}, \tag{32}$$

The *total* transmission probability for all the channels being given by

$$T = \sum_n T_n(E_n^\pm). \tag{33}$$

Similarly, in the case of the *opaque* barrier, the corresponding equations can be obtained from (11).

**4. Quantum Time of Barrier Traversal**

In Figure 4, we consider a time-modulated electron, with energy $E_N$, incident on a dynamic barrier at $x = -b/2$ and time $t_n$. On entering the barrier's stationary *elastic* channel *N*, whose energy becomes $E_N$, it is immediately subjected to the barrier's oscillating potential field, which *instantaneously* scatters it, inelastically, into the *sideband* time-quantized scattering channel +n (-n), of energy $E_n^\pm$ (23), by the *absorption* (*emission*) of photon energy ($\alpha = \hbar\omega$). At this stage, the electron travels along the +*n* (-*n*) scattering channel, until it is further instantaneously, inelastically scattered by the field into the time-quantized sideband scattering channel +*m* (-*m*), of energy $E_m^{\pm}$ ($\gtrless E_n^{\pm}$), again by the absorption (emission) of photon energy, in which it completes its passage across the barrier, and *exits* from it at $x = b/2$ and time $t_m$.

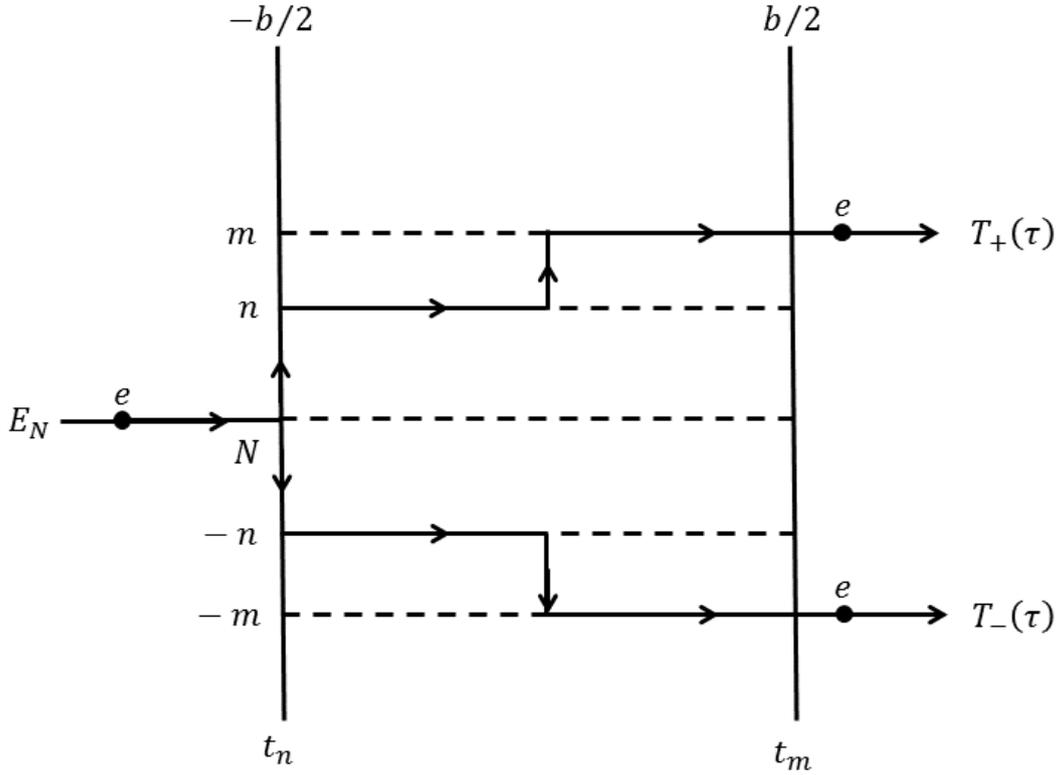

Figure 4. Schematic scattering diagram of electron (e) traversing dynamic barrier, depicting passage via positive (negative) scattering channels for photon absorption (emission) process, in low-frequency regime ($\omega T \ll 1$).

It should be mentioned that, in the above processes of electron absorption and emission of photons, which exist concomitantly, during the rapid barrier oscillations, the former (latter) occurs when the barrier is executing the upward (downward) part of its vibratory motion, and the electron is travelling among energy levels with greater (lesser) energy than itself.

By writing (19) in the form

$$\cos \omega t = n\omega\tau, \quad n = 0, \pm 1, \pm 2, \ldots, \tag{34}$$

where

$$\tau = h/V_1 = h/N\alpha = (N\omega)^{-1}, \tag{35}$$

we see that (34) represents the *quantization of time t* in units of *quantum time* $\tau$. By analogy with the photon of energy $\alpha\ (=\hbar\omega)$, equation (35) suggests that it is appropriate to refer to $\tau$ as a *timeon* of time, which can be viewed as a separate, discrete packet of time, dependent on the inverse amount of energy ($N\alpha$) contained in the corresponding potential barrier amplitude $V_1$ or the light frequency $\omega$. During the electron's transit of the barrier, the time scattering events are governed by (34), so that, at the electron's time of *entry* $t_n$ into the barrier, the $\pm n$ scattering channels are utilized. At this point, (34) transforms the continuous time into the discrete time $n\tau$. In the case of the barrier's *exit* time $t_m$, via the $\pm m$ scattering channels, (34) reads

$$\cos\omega t_m = m\omega\tau, \quad m < n, \tag{36}$$

where

$$t_m = t_n + T, \tag{37}$$

$T$ being the *time of barrier traversal* we seek. The plus (minus) sign in the above equations represents an absorption (emission) of photon energy. In the following calculations, it is mathematically convenient to treat separately the low (high) frequency regime of $\omega T \ll 1$ ($\omega T \gg 1$) in obtaining the $T(\tau)$ relations for the times of traversal.

## A. Low-Frequency Regime ($\omega T \ll 1$)

We begin, by using (34) and (37), to write (36) as

$$n\omega\tau \cos\omega T - [1 - (n\omega\tau)^2]^{1/2} \sin\omega T = m\omega\tau, \tag{38}$$

which, since $\omega T \ll 1$, can be taken to $O(\omega T)$, to obtain

$$n\omega\tau - [1 - (n\omega\tau)^2]^{1/2}\omega T \simeq m\omega\tau, \tag{39}$$

that leads to the *barrier traversal time* relation

$$T_\pm(\tau) \simeq \frac{\pm|n-m|N\tau}{(N+n)^{1/2}(N-n)^{1/2}}\ secs, \quad n > m, \tag{40}$$

on noting that

$$N\omega\tau = 1, \tag{41}$$

via (22) and (35).

Equation (40) is the scattering channel representation of the dynamic barrier time of traversal $T_\pm(\tau)$ in terms of the quantum of time $\tau$. The numerator contains the elastic channel $N$, together with the $\pm|n-m|$ term denoting the electron *transition* between the intermediate $n$ and the exit $m$ scattering channels for the absorption ($+$) and emission ($-$) of photon energy, the timeon $\tau$ being the unit of time in seconds. In the case of the denominator, the product of $(N + n)^{1/2}(N - n)^{1/2}$ describes the scattering process between the entrance elastic channel $N$ and the $+n$ ($-n$) scattering channel, during the absorption (emission) process. The *equality* of the absorption and emission times of traversal ($T_+ = T_-$) in (40) is in accord with the *same* finding of Lefebvre [16] for the transmission probabilities, when $\omega T \ll 1$. We note that, since $N \gg n$ in (40), a binomial expansion to $O(n^2/N^2)$ leads to the further approximate form of

$$T_\pm(\tau) \simeq \pm|n - m|(1 + n^2/2N^2)\tau. \tag{42}$$

Alternatively, using (23) enables (40) to be rewritten in terms of the energy levels, whereby we find that

$$T_\pm(E_n^\pm) \simeq \frac{\pm|n-m|h}{E_n^\pm - E_N} \text{ secs, } n > m, \tag{43}$$

via (41).

**B. High-Frequency Regime ($\omega T \gg 1$)**

This regime presents a somewhat more difficult problem than that of the low-frequency regime, because the condition of $\omega T \gg 1$ removes the simplifying features of the former case.

Setting $\theta = \omega T$, we write (38) as a quadratic equation, viz.,

$$A \tan^2 \theta_\pm + 2B \tan \theta_\pm + C = 0, \tag{44}$$

where

$$A = N^2 - n^2 - m^2, \quad B = n(N^2 - n^2)^{1/2}, \quad C = n^2 - m^2. \tag{45}$$

The roots of (44) are given by

$$\tan \theta_\pm = \frac{-n(N^2-n^2)^{1/2} \pm m(N^2-m^2)^{1/2}}{N^2-n^2-m^2}. \tag{46}$$

With the aid of (46), we can write

$$\frac{\tan \omega T_+}{\tan \omega T_-} = \frac{n(N^2-n^2)^{1/2} - m(N^2-m^2)^{1/2}}{n(N^2-n^2)^{1/2} + m(N^2-m^2)^{1/2}} < 1, \tag{47}$$

whereby we have

$$\sin \omega(T_+ - T_-) < 0, \tag{48}$$

whence,

$$T_+ < T_-. \tag{49}$$

Hence, the barrier traversal time for absorption is *shorter* than that for emission in this regime, which is also in keeping with the absorption and emission transmission probabilities for the $\omega T \gg 1$ in [16].

Having established the inelastic-scattering channel forms of the transmission probability (32), together with the low- and high-frequency times of dynamic-barrier traversal (40) and (46), respectively, further calculations are underway to perform a numerical analysis of these findings. The results of this work will appear in a future article.

## 5. Conclusion

An *exact* approach has been developed for studying the electron transmission of a time-modulated potential barrier. The crux of this approach lies in the *full* exploitation of the matching of the temporal parts of the wave-functions belonging to the free-space and dynamic barrier regions, which has hitherto been overlooked. Pursuing this line of enquiry, led to the crucial equation (19), which contains the essential source of the quantization of the scattering potential barrier field and its corresponding concomitant energy-level spectrum, and ultimately provided the important quantization of time, itself. The barrier-oscillation quantization enabled them to be represented by a set of successive snapshots of a sequence of static barriers of differing heights. Accordingly, only the electron energy of the wave-function is modified by the barrier-modulation process, while the functional form of the wave-function remains that of a static barrier. In other words, an electron in a scattering channel of a dynamic barrier, behaves as though it were in the elastic channel of an *equivalent* static barrier, whose elastic energy level has been modified by taking account of the time-modulation of the dynamic barrier. Turning to the energy spectrum itself, a *finite* number of discrete levels was found, whose band-width depended on the *size* of the modulation amplitude. Since the modulation amplitude is involved in generating these levels, such a result is not unexpected. However, it is at variance with the *infinite* spectrum of the BL theory, which, being a perturbation treatment, is limited to *small* modulation amplitudes. It is interesting to note that, in solving time-periodic problems, it is not always necessary to invoke the TDSE, since the alternative equivalent static-barrier route, taken here, is much simpler and direct, as witnessed in deriving the expression for the transmission probability of a dynamic barrier.

Solutions have been provided for the much sought quantum times of barrier traversal in the low- and high-frequency regimes, for both the absorption and emission of photons, which are in agreement with the findings of Lefebvre, in the case of transmission probabilities. In closing, we note that the replacing of the BL infinite energy spectrum by a finite one, restores *causality* to dynamic barrier tunnelling, which is in keeping with Einstein's "golden rule".

## 6. Acknowledgements

The authors wish to thank A.T. Amos and B.L. Burrows for interesting discussions during the early stages of this work. They are particularly grateful to M. Wagner for reading a preliminary version of the manuscript and providing many helpful comments. The authors are also indebted to P.G. Davison for her kind assistance in preparing the initial drafts of the manuscript.

## 7. Keywords

Electron tunnelling, oscillating barriers, time-traversal quantization, semi-empirical calculations

**References**

1. P.K. Tien and J.P. Gordon, Phys. Rev. **129**, 647 (1963).
2. J.H. Shirley, Phys. Rev. **138**, 979 (1965).
3. M. Büttiker and R. Landauer, Phys. Rev. Lett. **49**, 1739 (1982); Phys. Scr. **32**, 429 (1985).
4. M. Büttiker, Phys. Rev. B. **27**, 6178 (1983).
5. D Sokolovski and L.M. Baskin, Phys. Rev. A. **36**, 4604 (1987).
6. J. A. Stovneng and E. H. Hauge, J. Stat. Phys. **57**, 841 (1989).
7. M. Büttiker and R. Landauer, J. Stat. Phys., **58**, 371 (1990).
8. D. Sokolovski and J.N.L Connor, Phys Rev. A. **42**, 6512 (1990); W.S. Truscott, Phys. Rev. Lett. **70**, 1990 (1993); R. Landauer and Th. Martin, Rev. Mod. Phys. **66**, 217 (1994).
9. M. Wagner, Phys. Rev. B. **49**, 16544 (1994).
10. M. Wagner, Phys. Rev. Lett. **76**, 4010 (1996).
11. F. Grossmann, T. Dittrich, P. Jung and P. Hänggi, Phys. Rev. Lett. **67**, 516 (1991).
12. D. S. Saraga and M. Sassoli de Bianchi, Helv. Phys. Acta **70**, 751 (1997).
13. W. Li and L.E. Reichl, Phys. Rev. B. **60**, 15732 (1999).
14. D.F. Martinez and L.E. Reichl, Phys. Rev. B. **64**, 245315 (2001).
15. E. Merzbacher, *Quantum Mechanics*, 2ed. Wiley, New York, 1970.
16. R. Lefebvre, Theochem. **493**, 117 (1999).